\documentclass[twocolumn,prl,showpacs]{revtex4}

\usepackage{amsfonts,amsmath,amssymb,amsthm}
\usepackage{graphicx}

\newcommand{\lmax}{\lambda_{\rm max}}
\newcommand{\om}{\omega}

\newcommand{\Aff}{A_{\rm ff}}
\newcommand{\Afb}{A_{\rm fb}}

\newcommand{\eps}{\epsilon}
\newcommand{\aff}{a_{\rm ff}}
\newcommand{\afb}{a_{\rm fb}}

\begin{document}

\title{Reliability of Layered Neural Oscillator Networks}

\pacs{87.19.lj, 05.45.Xt, 05.45.-a\vspace*{-1ex}}

\author{Kevin K. Lin}
\affiliation{Department of Mathematics, University of Arizona\vspace*{-2ex}}
\email{klin@math.arizona.edu}
\author{Eric Shea-Brown}
\affiliation{Department of Applied Mathematics, University of Washington\vspace*{-2ex}}
\email{etsb@amath.washington.edu}
\author{Lai-Sang Young}
\affiliation{Courant Institute of Mathematical Sciences, New York University}
\email{lsy@cims.nyu.edu}
\date{May 22, 2008}

\begin{abstract}
We study the \textit{reliability} of large networks of coupled
neural oscillators in response to fluctuating stimuli.
Reliability means that a stimulus elicits essentially identical
responses upon repeated presentations.  We view the problem on
two scales: {\em neuronal reliability}, which concerns the
repeatability of spike times of individual neurons embedded
within a network, and {\em pooled-response reliability}, which
addresses the repeatability of the total synaptic output from
the network. We find that individual embedded neurons can be
reliable or unreliable depending on network conditions, whereas
pooled responses of sufficiently large networks are mostly
reliable. We study also the effects of noise, and find that some
types affect reliability more seriously than others.
\end{abstract}

\maketitle

The replicability of a system's response to external stimuli has
practical implications.  For example, if a sensory stimulus is
presented to a neural network multiple times, how similar are
the spike trains that it evokes?  The answer to this question,
{\em i.e.}, the {\em reliability} of the system, impacts the
precision of neural codes based on temporal patterns of
spikes~\cite{coding}.  Reliability issues are important in the
biological sciences, in optics, and in electronic circuit
theory.

This Letter discusses the reliability of networks in the context
of neuroscience, where a number of studies have been conducted
via analysis, simulations, and laboratory experiments. To
summarize, there is strong evidence that single neurons are
typically reliable~\cite{onecellexp,onecellphase,onecellother}.
However, for neurons embedded within large networks, a range of
behavior from reliable to unreliable is
seen~\cite{networkrel1,networkrel2,teramae,networkrelexp}.

From a theoretical standpoint, under what conditions is a
network reliable?  We answer this question for a class of neural
oscillator networks that are idealized models of commonly
occurring situations in neuroscience, namely networks with {\it
  layers}~\cite{layered}.  Specifically, we consider networks
with either one or two layers, with sparse intra-layer and
inter-layer connections.  Reliability of individual neurons and
their pooled responses are studied.  To make transparent the
mechanisms involved, we first neglect the effects of noise,
introducing it only later on.

The setup above can be seen as a driven dynamical system.
Because we are interested in large networks, the accompanying
dynamical systems have many degrees of freedom, making a
statistical approach desirable.  For this reason, and to
describe rapidly fluctuating stimuli and noise, we have chosen
to cast the problem in the framework of {\it random dynamical
  systems theory}.  Our findings are based on a combination of
qualitative theory and numerical simulations.

\smallskip
\noindent {\bf I. Model details.} Individual neurons are modeled
as phase oscillators or ``Theta neurons''; this is a common
model for neurons in intrinsically active, ``mean-driven" firing
regimes~\cite{model,EI}. We study pulse-coupled networks
described by equations of the form
\begin{equation}
  \dot\theta_i = \omega_i + z(\theta_i)~\Big[\sum_{j \neq i}
    a_{ji}\ g(\theta_j) + \eps_i I(t)\Big]~,
\label{eq.input}
\end{equation}
$i=1, \cdots N$, where $N\gg 1$ (see {\em e.g.}~\cite{model}).
The variables $\theta_i$ are the states of the neurons, {\em
  i.e.}  they are angles parameterized by $[0,1]$ with periodic
boundary conditions.  The $\om_i$ are intrinsic frequencies, and
the $a_{ji}$ are synaptic coupling strengths, mediated by a
smooth function $g \ge 0$ with $\int_0^1g(\theta)\ d\theta = 1$
and $g(\theta) > 0$ for
$\theta\in[-\frac{1}{20},\frac{1}{20}]$~\cite{g}.  That is to
say, neuron $j$ ``spikes" when $\theta_j=0$, exciting or
inhibiting neuron $i$ depending on whether $a_{ji}$ is $>0$ or
$<0$ ($a_{ji}=0$ means neuron $i$ does not receive direct input
from neuron $j$). The phase response curve is given by
$z(\theta)=\frac{1}{2\pi}[1-\cos(2\pi\theta)]$, as for ``Type
I'' neurons.  The stimulus is represented by $I(t)$, which we
take to be a ``frozen'' or quenched white noise, {\it i.e.},
$I(t)\ dt = dW_t$ where $W_t$ is a realization of standard
Brownian motion; we have found that the addition of
low-frequency components to $I(t)$ does not substantially change
our results.

We now explain how the parameters $\omega_i, a_{ji}$ and
$\eps_i$ in Eq.~(\ref{eq.input}) are chosen.  In a reliability
study of a fixed network, these parameters remain frozen, as
does $I(t)$, and each {\it trial} corresponds to a
randomly-chosen initial condition in the
system defined by~(\ref{eq.input}).

To incorporate some of the heterogeneity that occurs
biologically, we assume a $20\%$ variability in $\omega_i$ and
in the $a_{ji}$.  Specifically, the $\omega_i$ are drawn
randomly and independently from the uniform distribution on the
interval $[0.9,1.1]$.  (The $a_{ji}$ are discussed below.)

We study two types of layered network structures:

\noindent{\em Single-layer networks.}  We set $\eps_i\equiv\eps$
for all $i$, so that all neurons receive the same input $I(t)$
at the same amplitude $\eps$. We assume a $20\%$ connectivity
with mean synaptic strength $a$, {\it i.e.}, each neuron receives input from $\kappa = 0.2~N$ other neurons (chosen randomly in simulations), and
the nonzero $a_{ji}$ are drawn independently and 
uniformly from $[0.9a,1.1a]$.  The two main network parameters are thus $\eps$ and $a$.

\noindent{\em Two-layer networks.}  We divide the neurons into
two groups of size $\frac{N}{2}$ each, referred to as Layer 1
and Layer 2.  We set $\eps_i\equiv\eps$ for all neurons $i$ in
Layer 1, and $\eps_i\equiv 0$ in Layer 2.  Each neuron receives
connections from $\kappa = 0.2~N$ other neurons, with
$\frac{\kappa}{2}$ from its own layer and $\frac{\kappa}{2}$
from the other layer.  Intra-layer connections within Layer 1
(resp. Layer 2) have mean strength $a_1$ (resp. $a_2)$.  For
inter-layer connections, Layer $1\to 2$ connections have mean
strength $\aff$, while Layer $2\to 1$ connections have mean
strength $\afb$.  (Here, ``ff" and ``fb" refer to
``feedforward'' and ``feedback''.)  Actual, heterogeneous
coupling constants are randomly chosen to lie within $1 \pm 0.1$
of their mean values as before.  The main system parameters here
are $\eps$, $a_1$, $a_2$, $\aff$, and $\afb$.

\begin{figure}
\begin{center}
  \hspace*{-2ex}\includegraphics{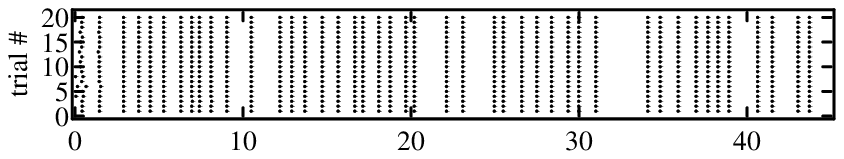}\\
  \vspace*{-2ex}
  $t$\\[1ex]
  \hspace*{-2ex}\includegraphics{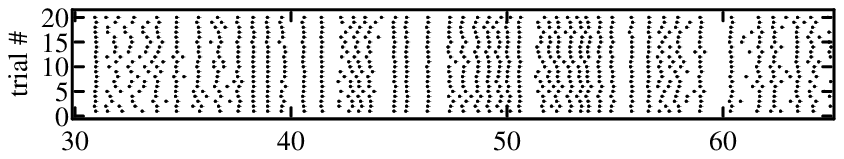}\\
  \vspace*{-2ex}
  $t$\\
\end{center}
\vspace*{-6ex}
\caption{Raster plots of single oscillators drawn randomly from
  two different networks.  Spike times are recorded for 20
  trials.  We set $\eps=2.5$ and $N=100$ in both numerical
  simulations.  {\em Top:} Single-layer model, $A=1$; $\lmax =
  -0.57$.  {\em Bottom:} Two-layer, $\Aff=2.8, \Afb=2.5,
  A_1=A_2=1$; $\lmax = 0.53$.}
\label{f.rasters}
\vspace*{-2ex}
\end{figure}

\smallskip
\noindent {\bf II. Neuronal reliability.}  This refers to the
repeatability of spike times from trial to trial for {\it
  individual neurons} within a network when the same stimulus
$I(t)$ is presented over multiple trials.  Fig.~\ref{f.rasters}
shows raster plots for two arbitrarily chosen neurons drawn from
two different networks.  The top panel shows repeatable spike
times; this is our definition of neuronal reliability.  The
bottom shows unreliability: spike times persistently differ from
trial to trial.  The latter cannot happen for single Theta
neurons {\it in isolation}, as they are always
reliable~\cite{onecellphase,onecellother}.

Neuronal reliability is closely related to stability properties
of the dynamical system defined by
Eq.~(\ref{eq.input})~\cite{onecellphase,onecellother,teramae,networkrel1,networkrel2}.
Recall that {\it Lyapunov exponents} measure the rates of
divergence of nearby orbits. These numbers make sense for
deterministic as well as random dynamical systems. For the
latter, under mild assumptions they are independent of initial
condition or realization of Brownian path (see~\cite{rds}).  Let
$\lmax$ denote the largest Lyapunov exponent of
(\ref{eq.input}).  The following are known mathematical
facts~\cite{LJ}: If $\lmax<0$, then regardless of the state of
the network at the onset of the stimulus, all trajectories
coalesce into a small region of phase space; this scenario,
referred to as a {\em random sink}, is equated with entrainment
to the stimulus and neuronal reliability.  Conversely, if $\lmax
>0$, the trajectories organize themselves around a complicated
object called a {\it random strange attractor}.  This means that
at a given point in time, the network may be in many different
states depending on its initial condition, {\it i.e.}, it is
unreliable.

\begin{figure}
  \begin{center}
    \begin{tabular}{cp{3ex}c}
      \includegraphics{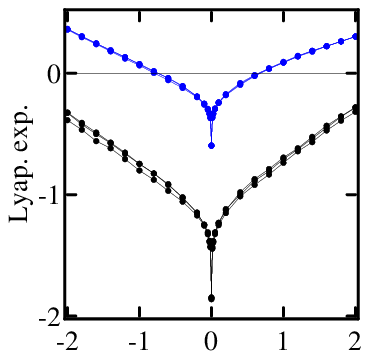}&&
      \includegraphics{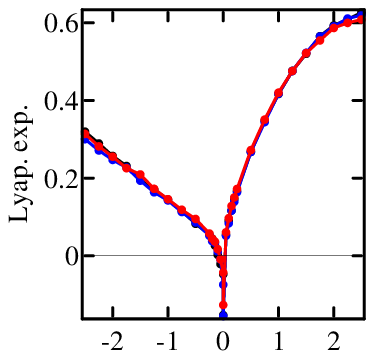}\\[-1ex]
      ~~~~$A$ && ~~~~~~~$\Afb$\\[1.5ex]
    \end{tabular}
  \end{center}
  \vspace*{-6ex}
  \caption{Lyapunov exponents $\lmax$.  {\em Left:}
    Single-layer, $N=100$, $\eps=1.5$ (top curve), $2.5$ (bottom curve). 
     {\em Right:}
    Two-layer, $N=100$, $\Aff=2.8, A_1= |A_2|=1$ (with
    $\mbox{sign}(A_2) = \mbox{sign}(\Afb)$), $\eps=2.5$.  Three
    realizations of network graphs are used in each case with their plots
    superimposed.}
  \label{f.single-layer-lyaps}
  \vspace*{-2ex}
\end{figure}

Our challenge here is to understand network reliability in terms
of the system parameters introduced above.  Measuring
reliability using a single quantity, $\lmax$, has the advantage
that large parts of the landscape can be seen at a glance, as in
Fig.~\ref{f.single-layer-lyaps}~\cite{lyap}.

\noindent{\em Single-layer networks.} We find that it is
fruitful to view $\lmax$ as a function of the quantity $A=\kappa
a$, which has the following interpretation: Focus on an arbitrary neuron,
say neuron $i$.  In the absence of any knowledge of the dynamics
({\em e.g.} firing rates), we expect each
of its $\kappa$ presynaptic neighbors to spike once per unit
time ($\om \approx 1$), with average strength $a$, and for
$z(\theta_i)$ to be at its mean value $\langle z \rangle =
\frac{1}{2 \pi}$, {\it i.e.}, we expect neuron $i$ to be
pushed (forwards if $A>0$ and backwards if $A<0$) by
$\frac{A}{2\pi}$ of a cycle per unit time.
If the dynamics are to approach a meaningful limit as
$N\to\infty$, it is necessary to stabilize the total synaptic
input received by a typical neuron.  Thus $A=\kappa a$ is a
natural scaling parameter.

\begin{figure}
\begin{center}
  \hspace*{\fill}%
  \resizebox{1.3in}{!}{\includegraphics{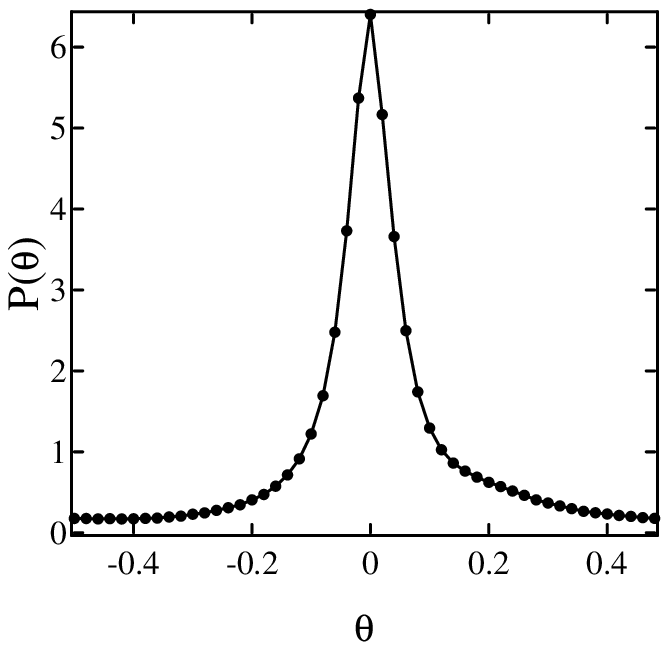}}\hspace*{\fill}%
  \resizebox{1.3in}{!}{\includegraphics{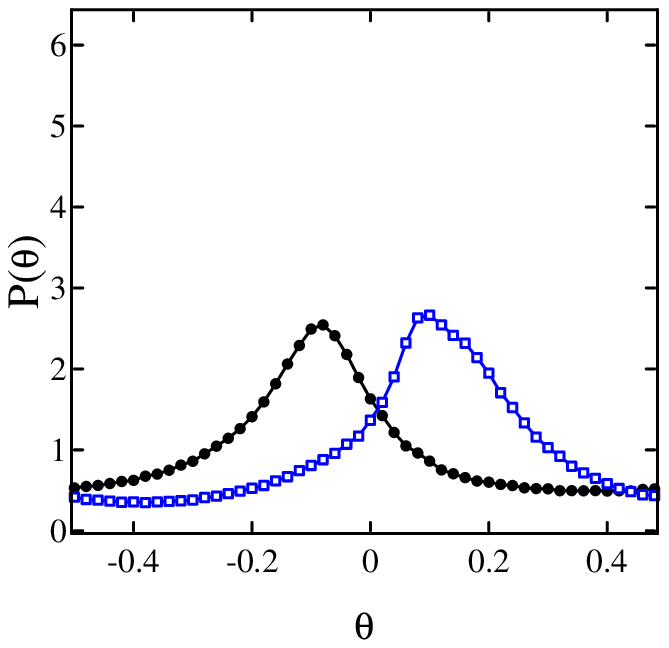}}\hspace*{\fill}
\end{center}
\vspace*{-6ex}
\caption{Phase distributions of neurons at the instant they receive an {\em incoming} spike.  {\em Left:} Single-layer, $A=1.8$; all spikes.
  {\em Right:} Two-layer, $\Aff=2.8$, $\Afb=0.8$, $A_1=A_2=1$; for
  inter-layer spikes only -- right peak for phases of Layer $1$ neurons, left peak for Layer $2$.}
\label{f.pdf}
\vspace*{-2ex}
\end{figure}

Fig.~\ref{f.single-layer-lyaps} (left) shows the basic
relationship between $\lmax, A$, and $\eps$ (stimulus
amplitude).  Plots for $1.5 < \eps < 2.5$ interpolate between
the two curves in a straightforward way.  When $A=0$, {\it
  i.e.}, when the oscillators are uncoupled, we have $\lmax <0$
as expected.  When $A\neq 0$, $\lmax$ can be positive or
negative.  Notice that (i) it increases with $|A|$ for fixed
$\eps$ (the sign of $A$ matters little), and (ii) it decreases
with $\eps$ for fixed $A$.  Item (ii) is due to the entraining
effects of the stimulus; (i) suggests that the couplings here
are intrinsically destabilizing.  We find the value of $\lmax$
to depend strongly on $A$, but only weakly on the underlying
balance of $N$, $\kappa$, and $a$ for large $N$.  Moreover,
$\lmax$ varies little among specific choices of connection graph
consistent with a given $\kappa$.

At first sight, the single-layer network may appear unexpectedly reliable:
At $A=2$, each neuron is expected to be perturbed
by $\frac13$ of a cycle per unit time, yet
Fig.~\ref{f.single-layer-lyaps} (left) shows $\lmax$ can still be negative.
This is due to the tendency of the network to synchronize, {\it
  i.e.}, to spike at roughly the same times (see
Fig.~\ref{f.pdf}, left).  Because $z(\theta)=z'(\theta)=0$ when
$\theta=0$, near-synchronization means that $z(\theta)$ is
typically quite small when a spike arrives, so that the {\em
  effective} total coupling strengths are considerably smaller
than the {\em a priori} strength $A$.  Perfect
synchrony is not possible here due to heterogeneity in the
$\omega_i$ and $a_{ji}$. For a given network
topology, greater homogeneity in $\omega_i$ and $a_{ji}$ will lead to
greater synchrony and smaller effective
coupling~\cite{paper3}.

\begin{figure}
\begin{center}
  \includegraphics{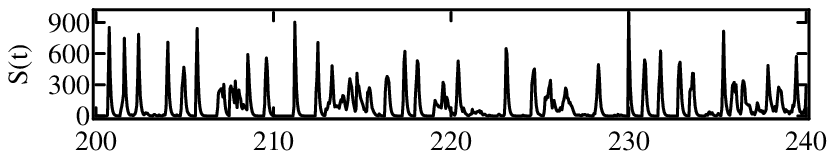}\\
  \includegraphics{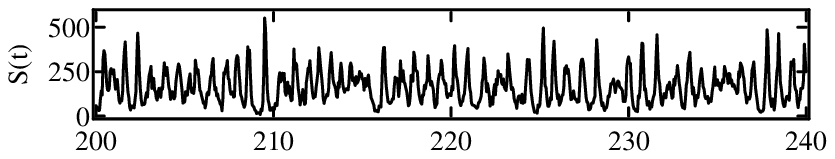}\\[-2ex]
  $t$
\end{center}
\vspace*{-5ex}
\caption{The bulk synaptic output function $S(t)$; one trial
  each for the two sets of parameters in Fig.~\ref{f.rasters}.}
\label{f.synaptic-output}
\vspace*{-2ex}
\end{figure}

\noindent{\em Two-layer networks.} We again express $\lmax$ in
terms of $A_1$, $A_2$, $\Aff$, and $\Afb$, defined to be
$\frac{\kappa}{2} = 0.1~N$ times $a_1$, $a_2$, $\aff$, and
$\afb$, respectively.  The interpretations are as before, {\it
  e.g.}, $\Aff$ is the {\it a priori} total kick received per
unit time by each neuron in Layer 2 from neurons in Layer 1.

Fig.~\ref{f.single-layer-lyaps} (right) shows $\lmax$ as a
function of $\Afb$ with $\Aff = 2.8$, $A_1=A_2= \pm 1$ (we give
$A_1$ and $A_2$ the same signs as $\Aff$ and $\Afb$,
respectively, as each neuron is either excitatory or
inhibitory), and $\eps =2.5$.  At $\Afb=0$, the system is
definitively reliable.  As $|\Afb|$ increases, however, we find that
the network loses its reliability almost immediately, even before $|\Afb| \approx \frac{1}{10}\Aff$.

This rather surprising fact is also partially explained by the
phase distributions of Layer 1 and Layer 2 neurons at the
instants when they receive inputs from the other layer (see
Fig.~\ref{f.pdf}, right).  The distributions are more spread out
than in the single-layer case; moreover, their peaks are
centered away from $\theta=0$. This can be predicted from
reduced two-neuron models~\cite{paper3}.  Thus at the same
numerical values, $\Aff$ and $\Afb$ in the two-layer model are
significantly more destabilizing than $A$ in the single-layer
model.  See also~\cite{feedback}.

We expect the ideas above, {\it i.e.}, the tendency to synchronize
within each layer, and the dominant effects of inter-layer interactions, 
to extend to multi-layer systems.

\begin{figure}
\begin{center}
  \hspace*{-3ex}\resizebox{!}{1.3in}{\includegraphics{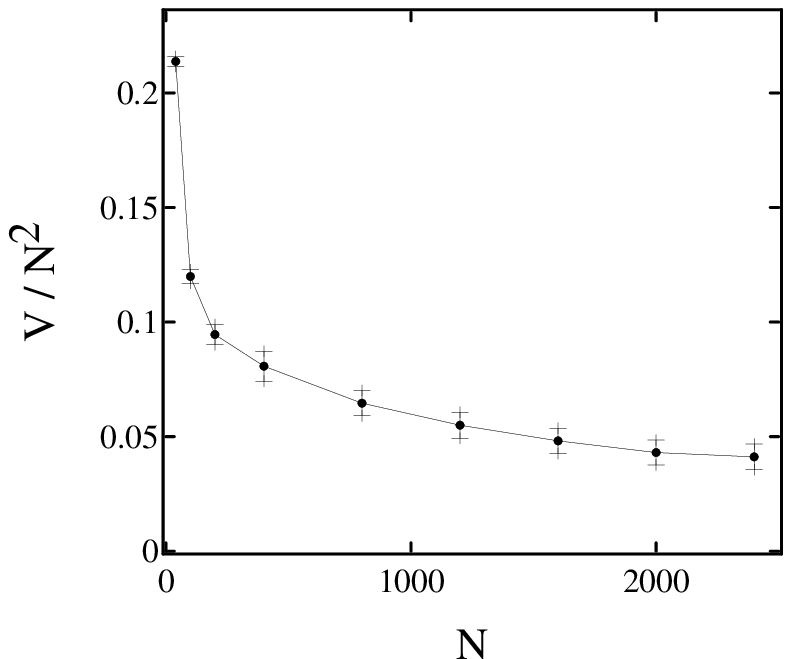}}~~~~%
  \resizebox{!}{1.3in}{\includegraphics{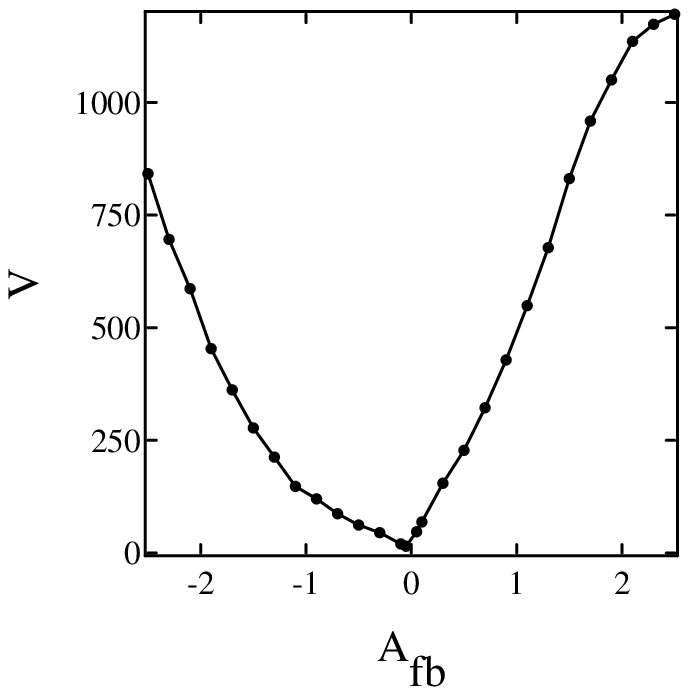}}\\
\end{center}
\vspace*{-6ex}
\caption{Mean across-trial variances $\bar V$. 
{\em Left:} $\bar V/N^2$ as function of $N$ for
  $\Aff=2.8, \Afb=2.5$. {\em Right:}  $\bar V$ as
  function of $\Afb$.  The setup is identical to that in
  Fig.~\ref{f.single-layer-lyaps} (right);
  $N=100$. }
\label{f.vbar}
\vspace*{-2ex}
\end{figure}

\smallskip
\noindent {\bf III. Reliability of pooled responses.}  Bulk
measurements have arguably greater impacts than the behavior
of individual neurons. A function representing the total synaptic
output of a system is defined by (cf.~\cite{maz,knight}) 
$$
S(t) = \sum f(t-T_i)~;\quad f(t) = \tau^{-1}e^{-t/\tau}~,
$$
where $T_i$ are spike times of {\em any} neuron in the
network, the summation is over all $T_i<t$,  $f$ is a postsynaptic current, 
and
$\tau\approx\frac{1}{15}$.  {\em Pooled-response reliability} describes
how repeatable $S(t)$ is in response to $I(t)$.
Clearly, neuronal reliability implies pooled-response reliability.
On the other hand, one would expect individual neurons to be more
volatile than the network as a whole.

Two time courses for $S(t)$ are shown in
Fig.~\ref{f.synaptic-output}.  The first is for a reliable
single-layer system; tall, well-defined spikes are generated
when the system is in partial synchrony.  The second is for an unreliable, 
2-layer model.  Here the floor of $S(t)$
is strictly positive, {\it i.e.}, some neurons in the system are
spiking at nearly all times.

For each $t$, we measure the repeatability of $S(t)$ by its
time-dependent, across-trial variance $V(t)$; this information
can be distilled further to give a single number $\bar V$ by
time averaging $V(t)$.  Our main finding is that $\bar V/N^2$
decreases as $N$ gets larger; see Fig.~\ref{f.vbar} (left).
Though beyond the reach of ergodic theorems, it is apparent that
due to the effects of averaging, total synaptic outputs of
sufficiently large networks tend to be reliable, even as
individual neurons behave unreliably.

Next, we fix $N$. As parameters are varied, we find strong
correlation between $\lmax$ and $\bar V$; compare
Figs.~\ref{f.single-layer-lyaps} (right) and \ref{f.vbar}
(right).  This confirms that the two different ways of measuring
unreliability we have proposed are in good qualitative
agreement.

\smallskip
\noindent
{\bf IV. Effects of noise.}  By ``noise", we refer to {\em
  trial-to-trial fluctuations} not modeled by
Eq.~(\ref{eq.input}).  For a clear conceptual understanding of its
impact on reliability, we find it useful  to
distinguish between (i) noise that affects each neuron
differently ({\it e.g.}  synaptic or membrane noise), and (ii)
noise that affects the entire population in essentially the same
way ({\em e.g.}, noise associated with the stimulus
$I(t)$)~\cite{maz,noise}.  As an idealization, we add to
Eq.~(\ref{eq.input}) two noise terms:
$$
  \dot\theta_i = \omega_i + z(\theta_i)\cdot \Big[\sum_{j \neq i}
    a_{ji}\ g(\theta_j) + \eps_i I(t) + \sigma_\ell \eta_i(t) + \sigma_g \zeta(t) \Big]
$$
Here $\zeta(t)$ and $\eta_i(t)$ are white noise realizations
which vary {\it independently from trial to trial};
additionally, the $\eta_i(t)$ are {\it independent for each $i$}.  
We refer to $\eta_i(t)$ and $\zeta(t)$ as ``local" and ``global" noise;
their respective amplitudes are denoted by $\sigma_\ell$
and $\sigma_g$.
 
Our simulations show that neuronal reliability persists under
some level of local and global noise (although a gradual
degradation of spike time precision from trial to trial is
unavoidable).  As expected, pooled responses can tolerate
higher-amplitude noise terms.

Since local noise is more varied, one might
expect it to lead to greater unreliability. This, however, is not true.
We find that local noise has only a limited
effect on pooled-response reliability for large networks, likely
due to averaging effects. In contrast,  the effects of global
noise can be much more severe.  The table below shows
$\bar V/N^2$ in two representative cases:
\vspace*{-1ex}
\begin{center}
  \begin{tabular}{lp{3ex}||p{3ex}ccc}
    &&&\multicolumn{3}{c}{Noise amp. $(\sigma_\ell, \sigma_g)$}\\[1pt]
    &&& (0, 0) &  (0.3, 0)   &   (0, 0.3)\\\hline
    Case 1: {\em Reliable} &&& 0.0 & 0.016 & 0.37 \\
    Case 2: {\em Unreliable} &&& 0.090  & 0.076 & 0.36  \\
  \end{tabular}
\end{center}
\vspace*{-1ex}
\noindent
Cases 1 and 2 are respectively the reliable and unreliable cases
in Fig.~\ref{f.rasters}.  The loss of reliability (in a
neuronally reliable system) due to global noise can be
understood as follows: Recall from Sec. II that reliability
means all trajectories independent of initial condition coalesce
into a ``random sink" for each $t$.  Within each trial, since
$\zeta(t)$ is a term of the same type as $I(t)$, its presence
strengthens the effects of the stimulus, leading to more robust
entrainment (see Fig.~\ref{f.single-layer-lyaps}, left).  Recall,
however, that
$\zeta(t)$ varies from trial to trial, so the trajectories
entrain to a {\it different} stimulus, and therefore coalesce to
a different state on each trial.  When $\sigma_g$ is large
enough, this provides a mechanism for destroying reliability.

\smallskip

\noindent {\bf Conclusion:} We have carried out a systematic
study of stimulus-response reliability for heterogeneous, layered
networks of neural oscillators.
Our findings -- all of which are new in the present context and
are consistent with results of earlier studies of different
models -- are of a very basic nature and thus are likely to shed
light on situations beyond those considered here:

\noindent (1) {\it On the neuronal level, single-layer networks
  are fairly reliable due to a tendency to synchronize, while
  recurrent connections can be strongly destabilizing in
  two-layer systems.}  In general, individual neurons can behave
reliably or unreliably as a result of the competition between
entrainment to the stimulus or upstream layer and the
perturbative effects of other synaptic events.

\noindent (2) {\it Pooled responses of large enough networks are
mostly reliable} even when individual
neurons within it are not. In a fixed-size network, they have
similar reliability properties as individual neurons but with
lower volatility.

\noindent (3) {\em Global noise, {\it i.e.}, noise that affects the entire
population in roughly the same way,  
can seriously jeopardize even pooled-response reliability}, 
while local noise has only limited effect.

\medskip
\begin{small}
E.S-B. is supported by a Burroughs-Wellcome Fund Career Award;
L-S.Y. is supported by a grant from the NSF. The authors thank
D. Cai and J. Rinzel for helpful discussions.
\end{small}


\begin{thebibliography}{42}
\bibitem{lyap} We compute $\lmax$ by solving the variational
  equation for the SDE~(\ref{eq.input}) using the Milstein
  scheme.
\end{thebibliography}
\end{document}